\documentclass{PoS}

\PoS{PoS(LAT2005)043}

\title{Light hadron spectroscopy in quenched QCD with overlap
fermions\thanks{This combined contribution includes content from
PoS(LAT2005)024.}.}

\ShortTitle{Light hadron spectroscopy in quenched QCD with overlap fermions. }

\author{Ronald~Babich$^a$, Federico~Berruto$^b$, Nicolas~Garron$^c$,
        Christian~Hoelbling$^d$, Joseph~Howard$^a$, Laurent~Lellouch$^e$,
        \speaker{Claudio~Rebbi}$\,^a$, and Noam~Shoresh$^f$\\
        \llap{$^a$}Department of Physics, Boston University, Boston, MA\\
        \llap{$^b$}Brookhaven National Laboratory, Upton, NY\\
        \llap{$^c$}DESY, Platanenallee 6, 15738 Zeuthen, Germany\\
        \llap{$^d$}Department of Physics, Bergische Universit\"at 
                   Wuppertal, Germany\\
        \llap{$^e$}Centre de Physique Th\'eorique\thanks{UMR 6207 du CNRS 
                   et des universit\'es d'Aix-Marseille I, II et du Sud
                   Toulon-Var, affili\'ee \`a la FRUMAM.}, Marseille, France\\
        \llap{$^f$}Harvard University, Cambridge, MA\\ \\
        E-mail: \email{rebbi@bu.edu}}

\abstract{A simulation of quenched QCD with the overlap Dirac operator
has been carried out using 100 Wilson gauge configurations at $\beta=6$
on an $18^3 \times 64$ lattice and at $\beta=5.85$ on a $14^3 \times 48$
lattice.  Here we present results for meson masses, meson final state
``wave functions,'' decay constants, and other observables, as well as details
on our algorithmic and data analysis techniques.  We also summarize results
for baryon masses and quark and diquark propagators in the Landau gauge.}

\FullConference{XXIIIrd International Symposium on Lattice Field Theory\\
		 25-30 July 2005\\
		 Trinity College, Dublin, Ireland}

\renewcommand{\logo}{
\raisebox{-5mm}{
\parbox{\textwidth}{
\includegraphics{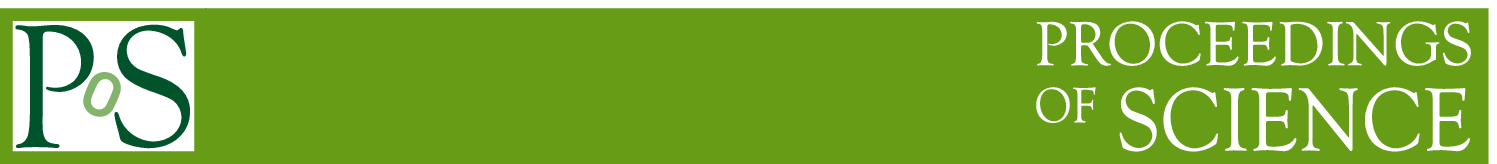}
\begin{flushright}
BUHEP-05-14 \\
CPT-2005/P.054
\end{flushright} }}}

\begin{document}

%%%%%%%%%%%%%%%%%%%%%%%%%%%%%%%%%%%%%%%%%%%%%%%%%%%%%%%%%%%%%%%%%%%%%%%%%%%%%%

\section{Introduction}

The importance of preserving chiral symmetry on the lattice and the advantages
of domain-wall \cite{Kaplan:1992bt,Shamir:1993zy} and overlap
\cite{Narayanan:1994gw,Narayanan:1993sk,Neuberger:1997fp} fermions are
well recognized, but these advantages come at a heavy computational cost.
It is therefore necessary to subject these formulations to the test of
simulations on lattices of realistically large size in order to explore the
adequacy of available numerical techniques and verify the benefits expected
to follow from the preservation of chiral symmetry.  In this contribution we
present the results of one such investigation with the overlap Dirac operator.

Quenched gauge configurations were generated with the Wilson gauge action
at $\beta=6$ on a lattice of size $18^3\times 64$, as well as at $\beta=5.85$
on a $14^3\times 48$ lattice in order to test scaling.  The corresponding
values of the lattice spacing are $a^{-1} = 2.12$~GeV and
$a^{-1} = 1.61$~GeV, on the basis of the Sommer scale defined by
$r_0^2F(r_0)=1.65$, $r_0=0.5$~fm~\cite{Guagnelli:1998ud}.  The two
lattices are therefore of approximately the same physical volume.
For 100 configurations at each of the two lattice sizes, overlap quark
propagators were calculated for a single point source and all color-spin
combinations using a conjugate gradient multimass solver, after gauge-fixing
to the Landau gauge.  Propagators were calculated for quark masses
$am_q=0.03,0.04,0.06,0.08,0.1,0.25,0.75$ on the $18^3\times 64$ lattice and
$am_q=0.03,0.04,0.053,0.08,0.106,0.132,0.33,0.66,0.99$ on the coarser
$14^3\times 48$ lattice.

Implementation of the overlap operator requires the calculation of
$H/\sqrt{H^\dagger H}$ where $H=\gamma_5(D_W-\rho/a)$, and $D_W$ is the Wilson
Dirac operator.  This was accomplished using a Zolotarev optimal rational
function approximation with 12 poles for the first 55 configurations on the
$18^3\times 64$ lattice.
%~\cite{vandenEshof:2002ms}.
A Chebyshev polynomial approximation
%~\cite{Bunk:1997wj,Hernandez:2000sb}
was used for all remaining
configurations after it was found to be about 20 percent more efficient.
In both cases, the lowest (12 for $18\times 64$, 40 for $14^3\times 48$)
eigenvectors of $H^2$ were first computed with a Ritz algorithm and projected
out before inverting the Dirac operator.
%~\cite{Kalkreuter:1995mm}.
As convergence criteria we required
$|1/\sqrt{H^2}-\sum T_n(H^2)|<10^{-8}$ and $|D^\dagger D\psi-\chi|<10^{-7}$.
The parameter $\rho$ in the definition of the overlap operator was chosen so
as to maximize locality and set to $\rho=1.4$ at $\beta=6$ and $\rho=1.6$
at $\beta=5.85$~\cite{Hernandez:1998et,Hernandez:1999cu}.  For further
details of the simulation, as well additional results and greater
discussion of the results presented herein, see~\cite{Babich:2005ay}.

In the sections that follow, we present results for the meson spectrum,
meson final state wave functions, the baryon spectrum, and quark and diquark
propagators calculated in the Landau gauge.

%%%%%%%%%%%%%%%%%%%%%%%%%%%%%%%%%%%%%%%%%%%%%%%%%%%%%%%%%%%%%%%%%%%%%%%%%%%%%%

\section{Light meson observables}

%%%%%%%%%%

\subsection{Pseudoscalar spectrum and quenched chiral logarithms}

In this and the following three sections, we present results obtained on the
finer lattice at $\beta=6$.  A comparison of the two lattices will follow in
Section~\ref{scaling-sec}.
We first consider meson correlation functions constructed with point sources
and sinks.  The general zero-momentum meson correlator is given by
\begin{equation}
G_{AB}(t) = \langle\,\sum_{\vec{x}} \mathrm{Tr}\left[
S^{f_2}(0;\vec{x},t)\,\Gamma_A\gamma_5\,(S^{f_1}(0;\vec{x},t))^\dagger
\,\gamma_5\Gamma_B\right]\rangle\;,
\end{equation}
where $S^{f_i}(0;\vec{x},t)$ is the Euclidean propagator for a quark of
flavor $f_i$, and $\Gamma_A$ and $\Gamma_B$ are the appropriate $\gamma$-matrix
combinations for the states of interest.  To extract ground state meson
masses, correlators were fit to the usual functional form
\begin{equation}
G(t) = \frac{Z}{M}\,e^{-MT/2}\cosh\left[M\left(\frac{T}{2}-t\right)\right]\;,
\label{eq-mesoncorrff}
\end{equation}
where $M$ is the meson mass, $T$ is the extent of the lattice in
time, and we refer to $Z$ as the correlator matrix element.  Fits were
performed within specified windows $t_{min}\le t \le 32a$, which were
determined by effective mass estimates and by scanning candidate values of
$t_{min}$ to find the smallest value (consistent with the errors) before
the clear effect of higher states caused the mass prediction to rise.
Errors in the meson masses were estimated with the bootstrap method with
300 samples.

\begin{figure}
\begin{minipage}[t]{0.48\textwidth}
\includegraphics*[width=\textwidth]{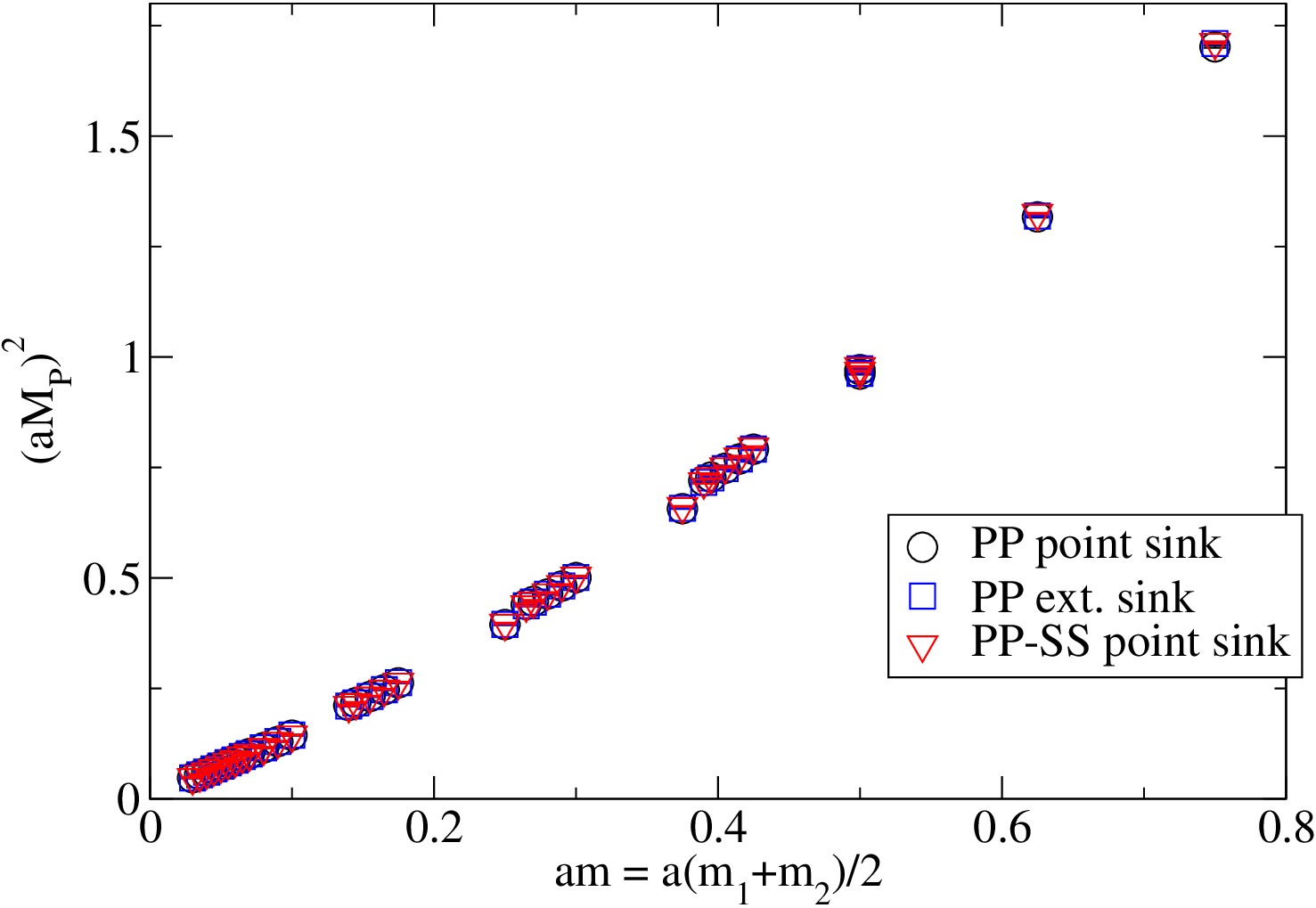}
\caption{\label{fig-PPall}Pseudoscalar meson spectrum for both point and
extended sinks.}
\end{minipage}
\hfill
\begin{minipage}[t]{0.48\textwidth}
\includegraphics*[width=\textwidth]{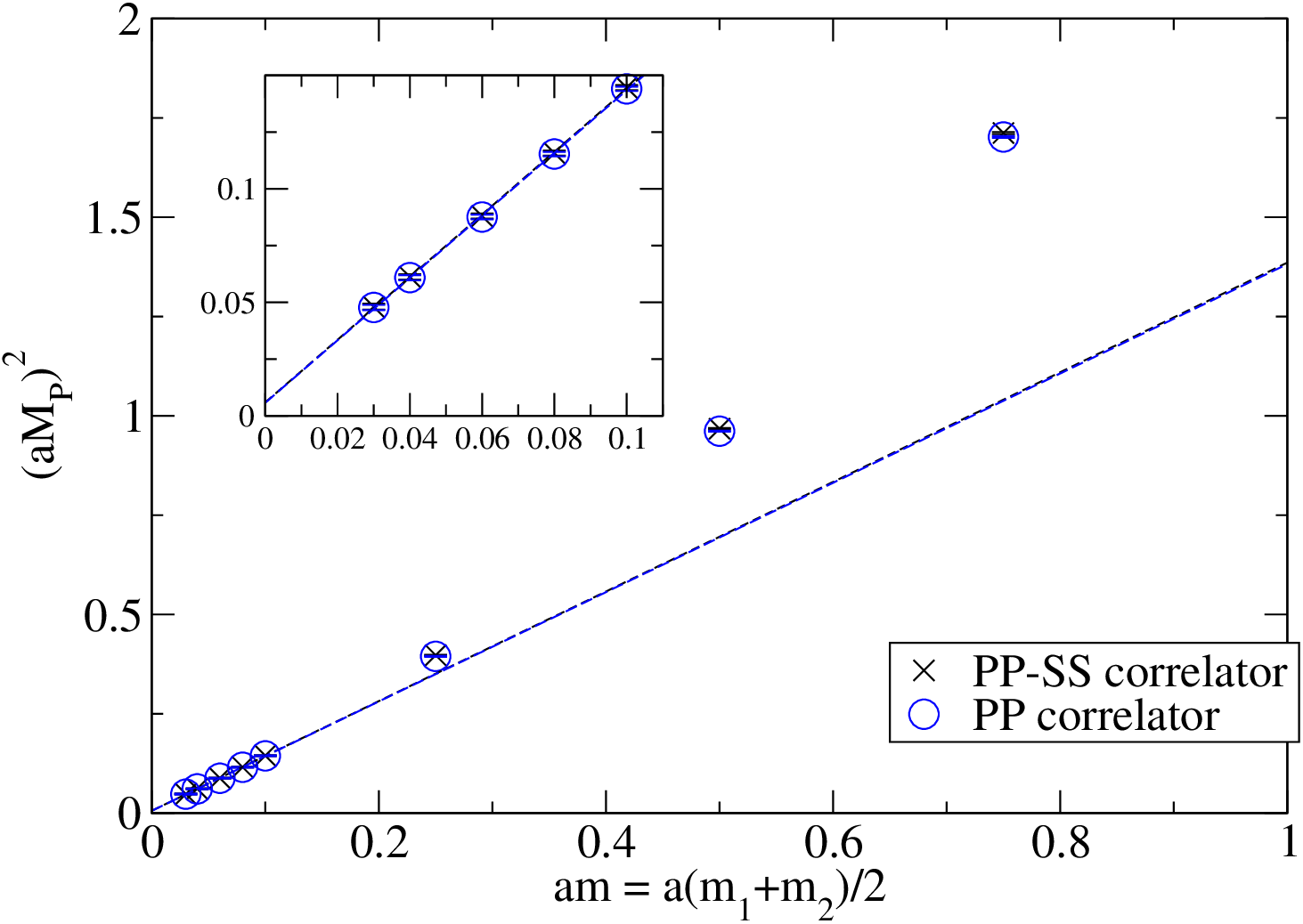}
\caption{\label{fig-mqmp2}Chiral fit of pseudoscalar meson masses.}
\end{minipage}
\end{figure}

We plot in Fig.~\ref{fig-PPall} our results for the pseudoscalar 
spectrum for all possible input quark mass combinations. The figure also
includes results for correlators with extended sinks, which will be discussed
in the next section. In the quenched approximation at finite volume, the
correlator $G_{PP}(t)$ receives contributions proportional to $1/m^2$ and
$1/m$ from chiral zero modes that are not suppressed by the fermionic
determinant.  These may be eliminated by considering
the difference of pseudoscalar and scalar meson correlators
$G_{PP-SS}(t) = G_{PP}(t) - G_{SS}(t)$,
since the quenching artifacts cancel by chirality in the
difference~\cite{Blum:2000kn}.  On our large lattice no
significant differences between the results obtained with $PP$ and $PP-SS$
correlators were observed.  Except where noted, we therefore make use of
$PP$ correlators in the remainder of this section.

Neglecting for the moment chiral logarithms, we fit the pseudoscalar
correlators to the linear form
\begin{equation}
(aM_P)^2={\cal A}+{\cal B}(am)\;,
\end{equation}
for quark masses $am\le 0.1$.  This yields
${\cal A}=0.0058(15),\ {\cal B}=1.376(15)$
for the $PP$ correlator and 
${\cal A}=0.0059(16),\ {\cal B}=1.380(17)$
for the $PP-SS$ correlator.
The nonzero value of the intercept ${\cal A}$ is statistically significant,
and the $PP-SS$ result confirms that it is not attributable to zero modes.
The deviation from linear behavior must therefore be due either to finite
volume effects or to chiral logarithms.  Regarding finite volume effects, we
can note that the Compton wavelength of our lightest pseudoscalar,
$M_P^{-1} \approx 4.5 a$, is much smaller than the lattice size $L=18a$
($M_PL=4$).

\begin{figure}
\begin{center}
\includegraphics*[width=0.5\textwidth]{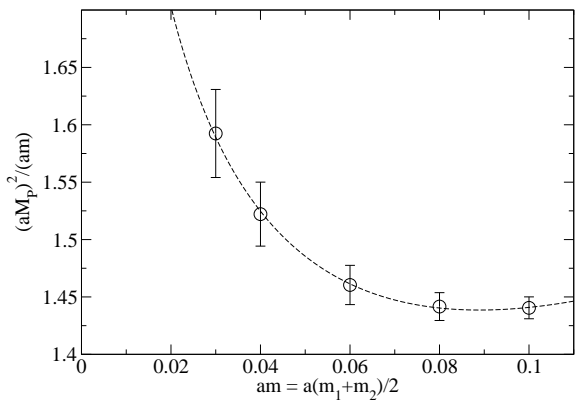}
\vspace{-5mm}
\end{center}
\caption{\label{fig-msq1-PP}Evidence for quenched chiral logs.}
\end{figure}

We proceed to test the supposition of quenched chiral logs, and fit the
degenerate quark mass results for the pseudoscalar masses to the
expression~\cite{Sharpe:1992ft}
\begin{equation}
(aM_P)^2=A(am)^{1/(1+\delta)}+B(am)^2\;,
\end{equation}
where the leading quenched logarithms have been resummed into a power
behavior, and the term proportional to $B$ parameterizes possible
higher-order corrections in the mass expansion.
The resulting fit is shown in Fig.~\ref{fig-msq1-PP}, plotted in terms of the
ratio $(aM_P)^2/am$, which exhibits a sharp rise at small $m$.  This yields
$A=0.680(68),\ B=2.98(31),\ \delta=0.29(5)$, consistent with values of
$\delta$ presented elsewhere in the literature.
%~\cite{Kim:1999ur,Blum:2000kn,
%DeGrand:2002gm,Aoki:2002fd,Draper:2002ep,Gattringer:2003qx,Chiu:2005ue}.

%%%%%%%%%%

\subsection{Extended sinks and vector meson spectrum}

For some correlation functions, such as the vector VV correlators, signals
may be greatly improved by the use of extended sources and sinks.  As
a practical constraint, we were forced to generate quark propagators for point
sources, as these were required for the study of nonperturbative 
renormalization and the evaluation of selected matrix elements.  We were
free to take advantage of extended sinks, however, which were constructed
for $PP$ and $VV$ correlators as follows.

\begin{figure}
\begin{minipage}[t]{0.48\textwidth}
\includegraphics*[width=\textwidth]{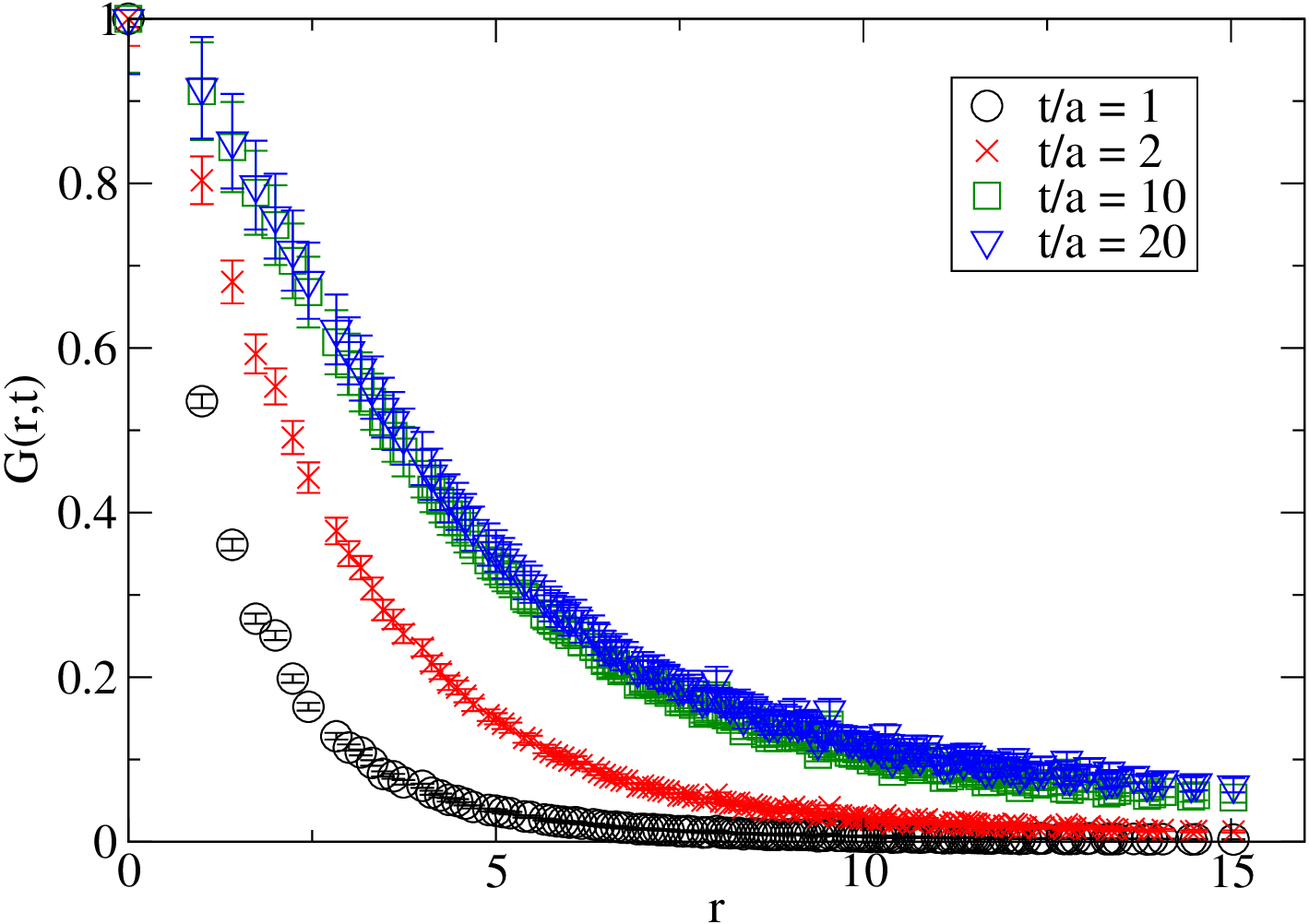}
\caption{\label{fig-PPGrt0303}$PP$ extended sink correlators $G(r,t)$ at
various $t$ for $am_1=am_2=0.03$.}
\end{minipage}
\hfill
\begin{minipage}[t]{0.48\textwidth}
\includegraphics*[width=\textwidth]{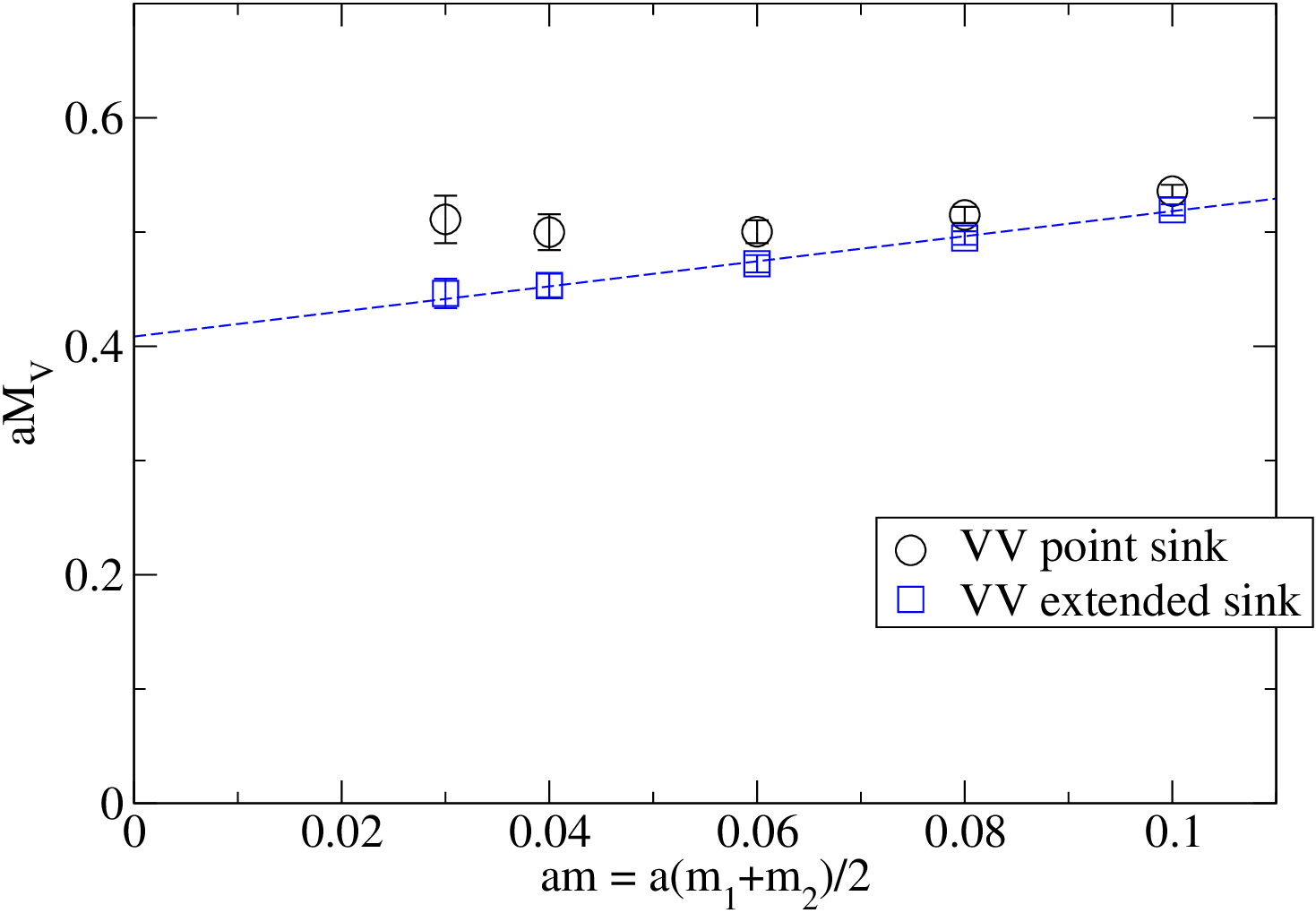}
\caption{\label{fig-VVlowmq}Vector meson spectrum for $am\le 0.1$.}
\end{minipage}
\end{figure}

In the Landau gauge, we define a correlation function dependent on the
separation $r$ between the quark and antiquark at the
sink~\cite{Gottlieb:1985xq} by
\begin{equation}
G(r,t)=\sum_{\vec{x},\vec{y}} \left\{ \mathrm{Tr}\left[S^{f_2}(0;\vec{x},t)\,
\Gamma_A\gamma_5\,(S^{f_1}(0;\vec{y}, t))^\dagger\,\gamma_5\Gamma_B\right]
\delta(\vert\vec{x} - \vec{y} \vert -r)\right\}\;.
\label{ext-corr}
\end{equation}
In calculating this function, a fast Fourier transform was used to
reduce the double summation over spatial lattice sites to a single sum,
decreasing the computational time by almost three orders of
magnitude~\cite{Hauswirth:2002eq}.  Figure~\ref{fig-PPGrt0303} shows the
mean value of the $G(r,t)$ correlators for pseudoscalar mesons with quark
mass $am=0.03$, each normalized to unity at $r=0$.  A clear ground state
``wave function'' is apparent after about $t/a=8$.  For both the $PP$
and $VV$ correlators, we used the corresponding functions
$\varphi(r) \equiv G(r,8a)$ to define extended sink correlators
\begin{equation}
\label{ext-sink-corr}
G_\mathrm{ext}(t)=\sum_r \varphi(r) G(r,t) \;,
\end{equation}
from which we then extracted the meson masses.

The use of extended sinks was most valuable in the calculation of the
vector meson spectrum.  The point sink and extended sink vector meson
spectra are compared in Fig.~\ref{fig-VVlowmq} for small quark masses
$am\le 0.1$.  A linear fit of the extended sink data gives
$aM_V=0.409(15)+1.10(13)\,(am)$.  We note that the chiral limit value of
0.409(15) is larger than the value 0.366 obtained using the lattice spacing
as determined by the Sommer scale and the experimental $\rho$ mass, giving an
indication of the systematic error induced by the quenched approximation.

%%%%%%%%%%

\subsection{Axial Ward identity and $Z_A$}

\begin{figure}
\begin{minipage}[t]{0.48\textwidth}
\includegraphics*[width=\textwidth]{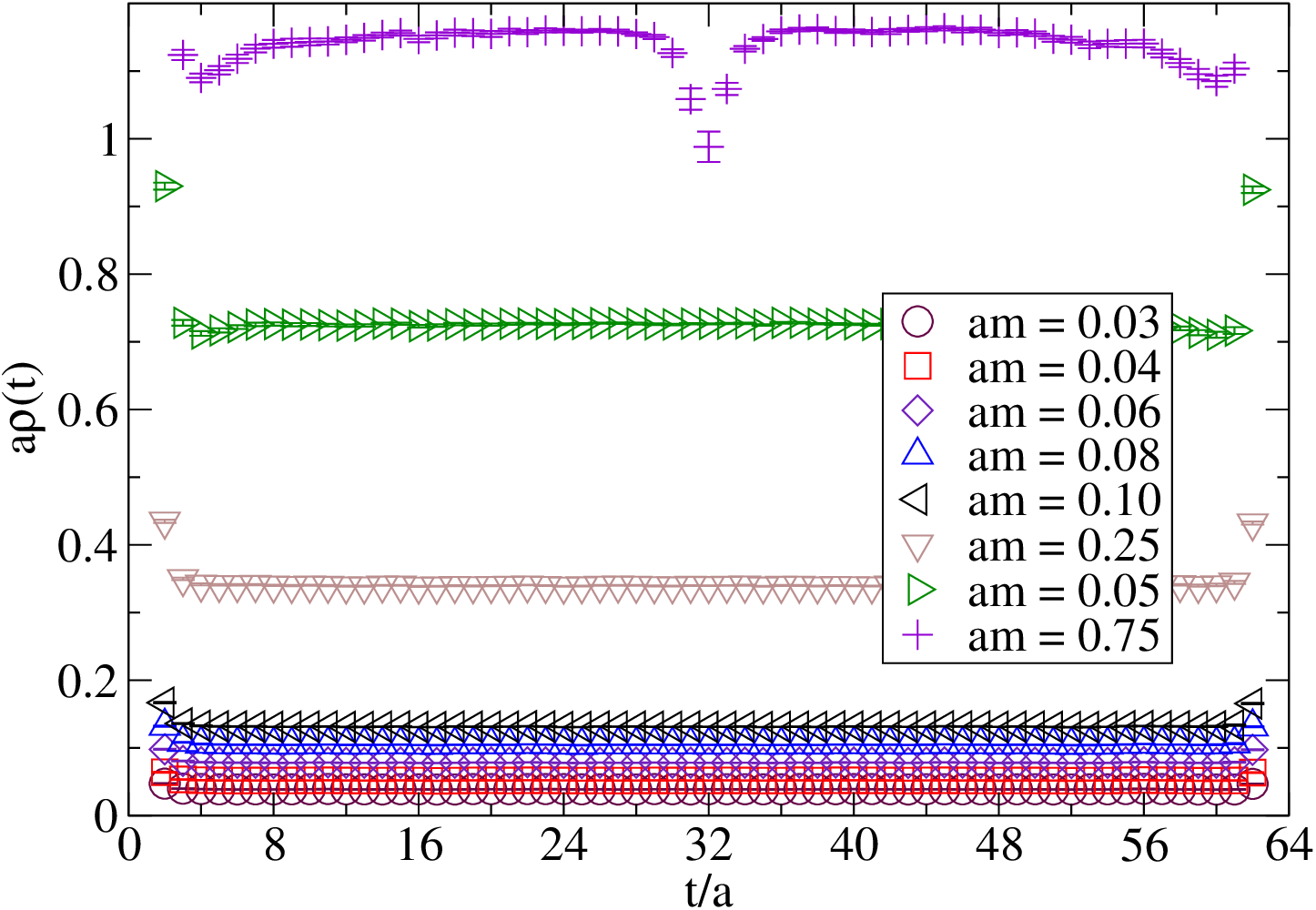}
\caption{\label{fig-rhot}AWI ratio as a function of time for all degenerate
quark mass combinations.}
\end{minipage}
\hfill
\begin{minipage}[t]{0.48\textwidth}
\includegraphics*[width=\textwidth]{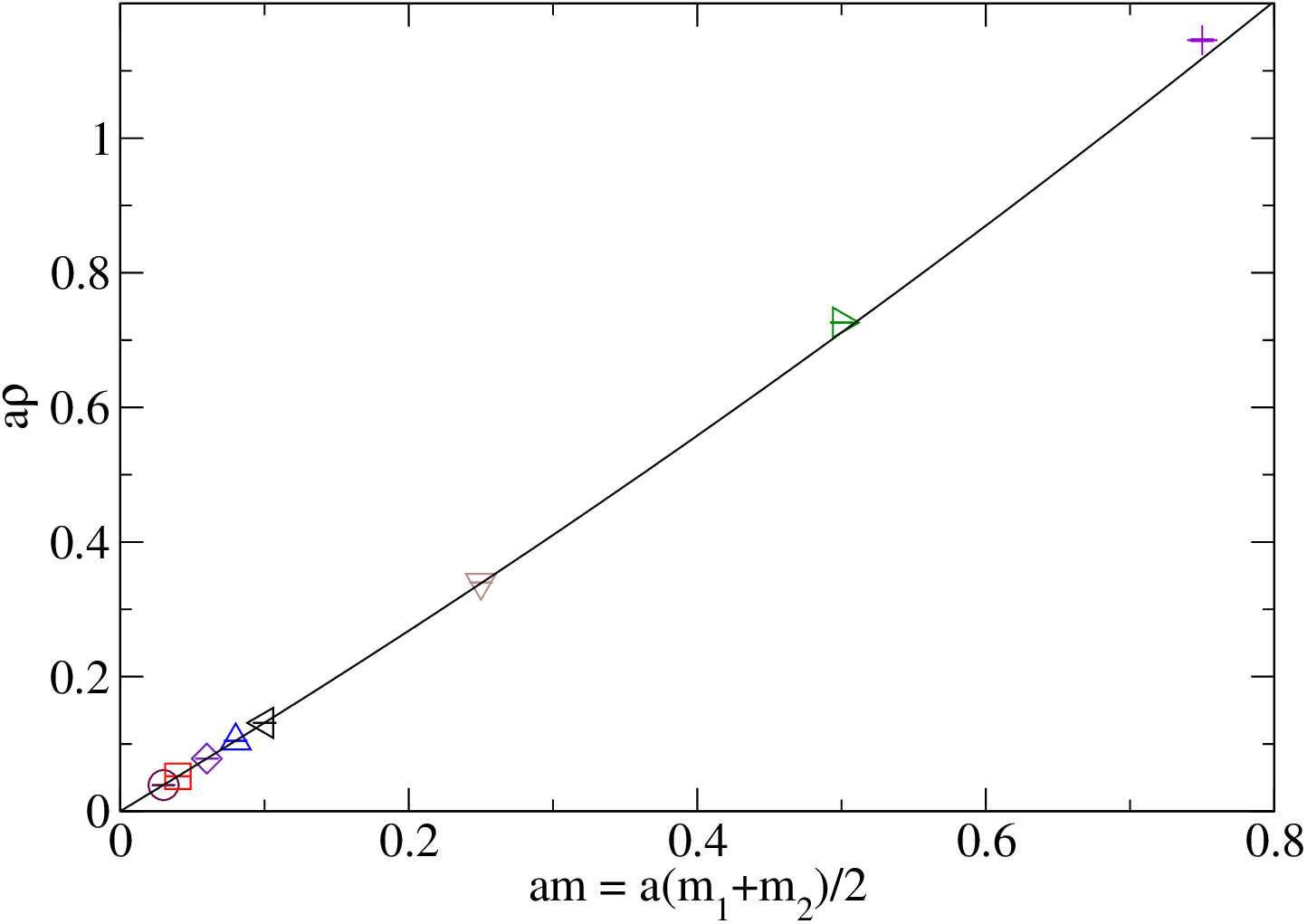}
\caption{\label{fig-rho}Axial Ward identity fit.}
\end{minipage}
\end{figure}

Exact chiral symmetry implies a conserved axial current, and the
associated axial Ward identity (AWI) predicts a constant value for the ratio
\begin{equation}
\label{eq-rho}
\rho(t)=\frac{G_{\nabla_0 A_0 P}(t)}{G_{P P}(t)}\;.
\end{equation}
The conserved axial current is a local, but not ultralocal, operator.
The ultralocal axial current 
\begin{equation}
\label{eq-axialc}
A_0=\bar\psi_1(x)\gamma_0\gamma_5\left[(1-\frac{a}{2\rho}D)\psi_2\right](x)
\end{equation}
differs from the exactly conserved axial current by a finite renormalization
factor $Z_A$ and possible corrections ${\cal O}(a^2)$.  We calculated
the correlator in Eq.~(\ref{eq-rho}) with the current of Eq.~(\ref{eq-axialc})
and using the lattice central difference for $\nabla_0$, corrected
so as to take into account the $\rm sinh$ behavior of the correlator.
Figure~\ref{fig-rhot} shows the observation of plateaus for all available
quark masses in the range $8 \le t/a \le 56$.

The fit shown in Fig.~\ref{fig-rho} to
\begin{equation}
a\rho={\cal A}+2(am)/Z_A+ {\cal C}(am)^2
\label{eq-AWIfit}
\end{equation}
gives ${\cal A} = 0.00002(10),\ Z_A = 1.5555(47),\ {\cal C} = 0.273(32)$.
The fact that ${\cal A}$ is consistent with zero is an excellent indication
of the good chiral behavior of the overlap formulation (compared to the
residual mass found in domain-wall fermion calculations).  We also note
that ${\cal C}$ is rather small, possibly indicating that discretization
errors might be smaller than expected on the basis of purely dimensional
arguments.

%%%%%%%%%%

\subsection{Quark masses and chiral condensate}

Using our data for the pseudoscalar spectrum together with the experimental
value for the kaon mass rescaled by the lattice spacing as determined from the
Sommer scale, we find $a(m_s+\hat{m})=0.0709(17)$ for the sum of the strange
and light bare quark masses.  A given bare quark mass $m(a)$ is related to the
renormalized quark mass $\bar m(\mu)$ by
\begin{equation}
\bar m(\mu)=\lim_{a \to 0} Z_m(a\mu)m(a)\;.
\end{equation}
The mass renormalization constant $Z_m$ is in turn related to the
renormalization constant $Z_S$ for the non-singlet scalar density by
$Z_m(a \mu)=1/Z_S(a \mu)$.  We calculated $Z_S$ in the RI-MOM scheme
starting from the identity
\begin{equation}
Z_S^{\mathrm{RI}}(a \mu)=\lim_{m \to 0}Z_A
\left.\frac{\Gamma_A(p,m)}{\Gamma_S(p,m)}
\right|_{p^2=\mu^2}\;,
\end{equation}
where $\Gamma_A(p,m)$ and $\Gamma_S(p,m)$ are suitably defined quark two-point
functions for the axial current and the scalar density in the Landau gauge, and
$Z_A$ is the renormalization constant for the axial current calculated in the
previous section.
Details of the procedure will be presented in a forthcoming publication.  
From it we find $Z_S^{\mathrm{RI}}(2\,\mathrm{GeV})=1.195(9)(27)$,
where the first error is statistical and the second systematic.  
With this result, one can use the three-loop perturbative calculation 
of the ratio $Z_S^{\overline{\mathrm{MS}}}/Z_S^{\mathrm{RI}}$ 
\cite{Chetyrkin:1999pq} to calculate
$Z_S^{\overline{\mathrm{MS}}}(2\,\mathrm{GeV})=1.399(10)(32)$.
Using this value, we find
$(m_s+\hat{m})^{\overline{\mathrm{MS}}}(2\,\mathrm{GeV})=107(4)(2)$~MeV
for the sum of strange and light quark masses.
Finally, using the value $m_s/\hat{m}=24.4(1.5)$ from chiral perturbation
theory~\cite{Leutwyler:1996qg}, we obtain
$m_s^{\overline{\mathrm{MS}}}(2\,\mathrm{GeV})=103(4)(2)$~MeV
for the strange quark mass.

We have also calculated the chiral condensate by performing a fit to the mass
dependence of the quantity
\begin{equation}
-a^3\tilde{\chi}(m)=am\sum_x(\langle P(x)P^c(0)\rangle
-\langle S(x)S^c(0)\rangle)\;,
\end{equation}
where the superscript $c$ denotes interchange of the two quark flavors.
A quadratic fit gives $-a^3\chi=0.00131(8)$ for the value at zero quark
mass, yielding $\langle\bar\psi \psi\rangle^{\overline{\mathrm{MS}}}
(2\,\mathrm{GeV})=-0.0175(11)(6)\,\mathrm{GeV}^3
= -[260(6)(8)\,\mathrm{MeV}]^3$.

%%%%%%%%%%

\subsection{\label{scaling-sec}Meson scaling}

\begin{figure}
\begin{minipage}[t]{0.48\textwidth}
\includegraphics*[width=\textwidth]{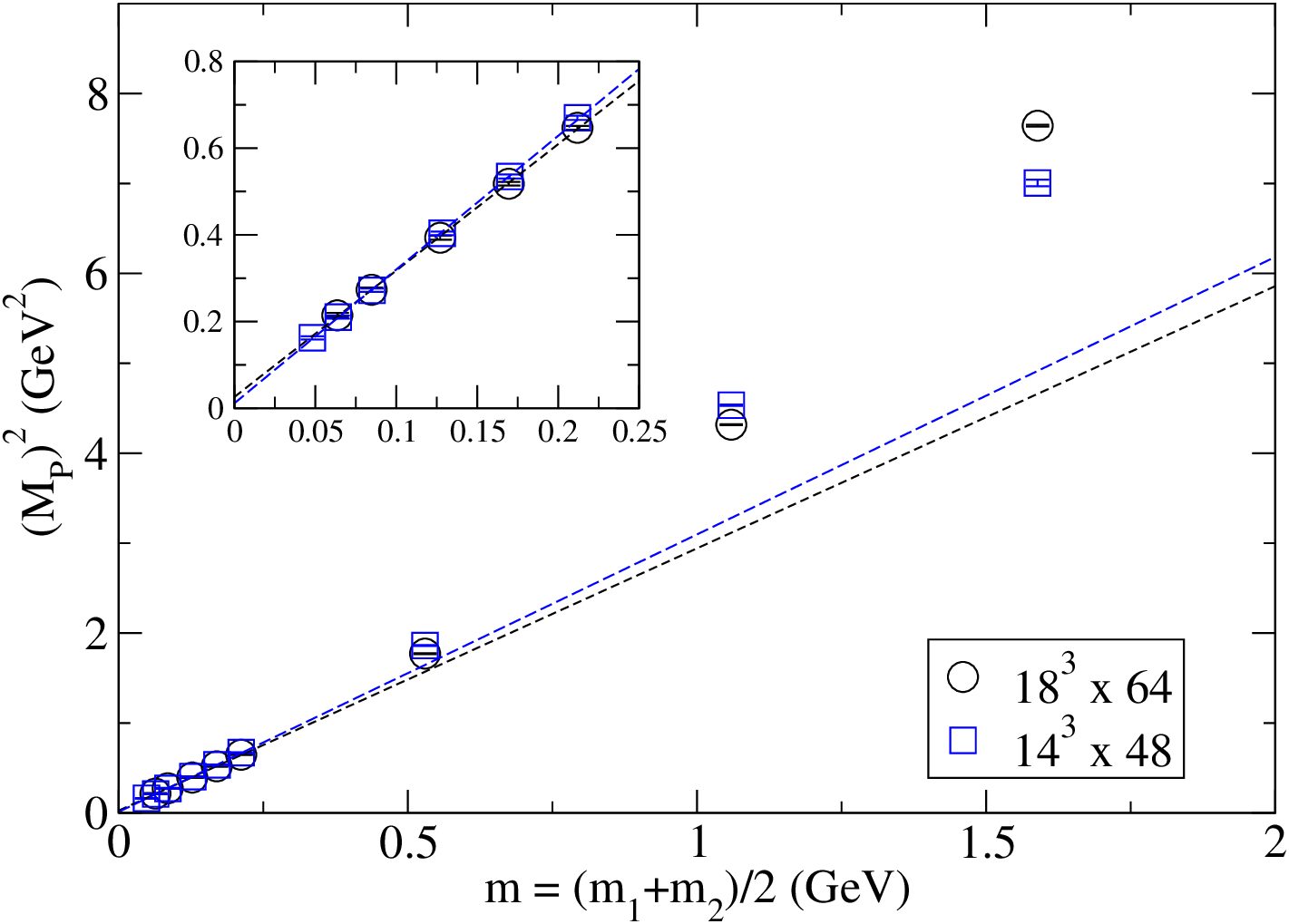}
\caption{\label{fig-MPSq_scaling_PP}Pseudoscalar $PP$ spectrum comparison.}
\end{minipage}
\hfill
\begin{minipage}[t]{0.48\textwidth}
\includegraphics*[width=\textwidth]{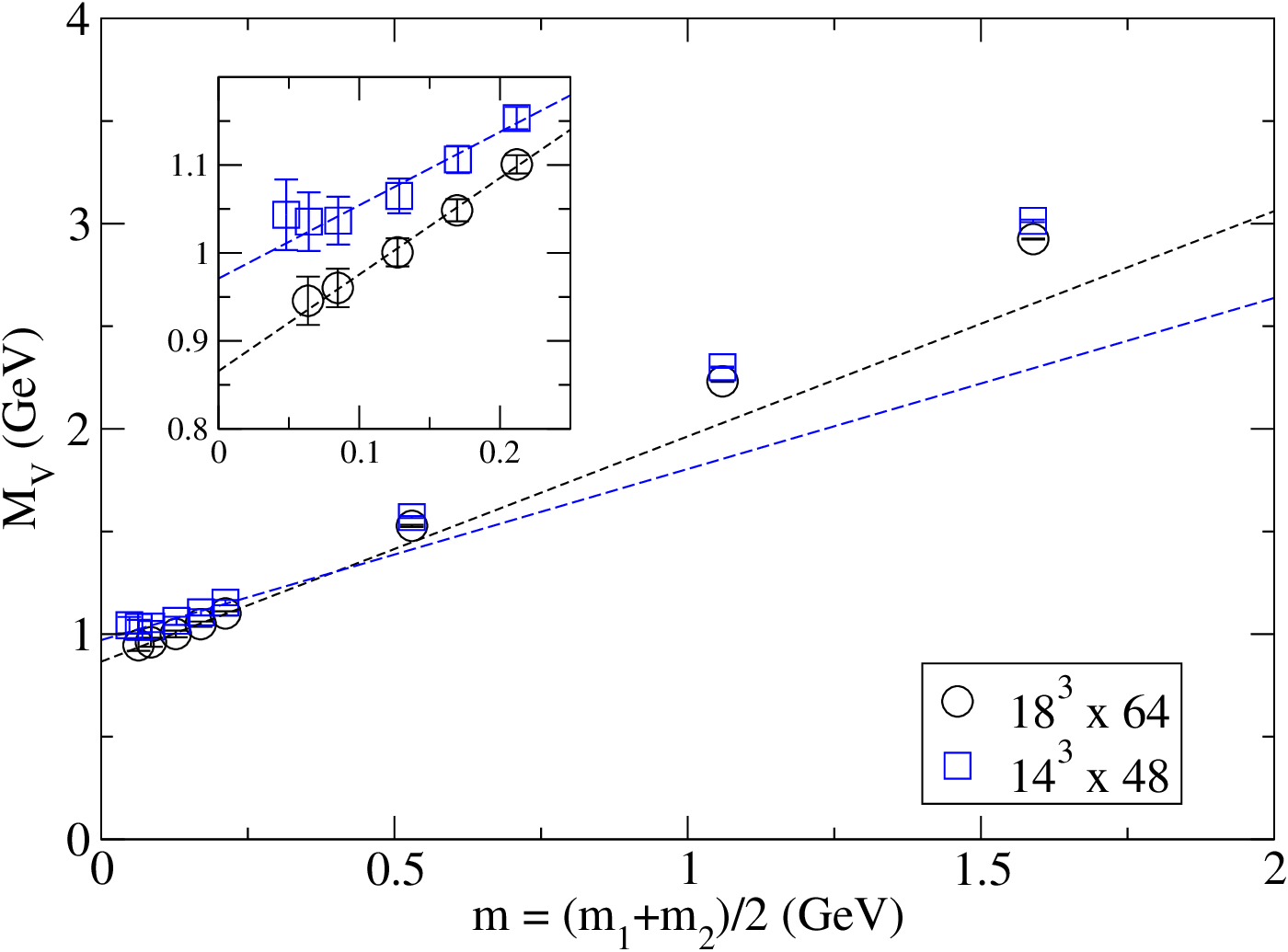}
\caption{\label{fig-MV_scaling}Vector (extended sink) spectrum comparison.}
\end{minipage}
\end{figure}

In Figs.~\ref{fig-MPSq_scaling_PP} and~\ref{fig-MV_scaling}, we compare our
results for the pseudoscalar and vector spectra on the two lattices,
using the lattice spacing determined from the Sommer scale to express masses
in physical units. We neglect logarithmic effects in the lattice spacing
and plot the mass spectra as a function of bare quark mass.  It is
interesting to observe that our results for the mass spectra on
the two different lattices are very similar, a conclusion that would remain
qualitatively unchanged in considering renormalized quark masses.  Our
results suggest that scaling violations for these quantities may be quite
small.

Complete results for the coarser lattice may be found in~\cite{Babich:2005ay},
and we provide a direct comparison of selected quantities in
Table~\ref{tab-scaling}.

\begin{table}
\begin{center}
\begin{tabular}{|l|l|l|}
\hline
Quantity & $\beta=6$ & $\beta=5.85$ \\
\hline \hline
$Z_A$ & 1.5555(47) & 1.4432(50) \\
$\delta$ (degenerate) & 0.29(5)  & 0.17(4)  \\
$a^{-1}$ (Sommer)~\cite{Guagnelli:1998ud} & 2.12~GeV  & 1.61~GeV  \\
$a^{-1}$ (physical planes) & $2.19(6)$~GeV  & $1.44(4)$~GeV  \\
$a^{-1}$ ($M_\rho$) & $1.90(4)$~GeV & $1.28(6)$~GeV \\
$f_K/f_\pi$ & 1.13(4)  & 1.09(4)  \\
$f_{K^*}/f_\rho$ & 1.03(6)  & 1.03(10)  \\
$M_{K^*}/M_\rho$ & 1.09(5) & 1.08(6)  \\
$Z_S^{\overline{\mathrm{MS}}}(2\,\mathrm{GeV})$ & 1.399(10)(32) &
     1.290(14)(82)  \\
$(m_s+\hat{m})^{\overline{\mathrm{MS}}}(2\,\mathrm{GeV})$ & $107(4)(2)$~MeV &
     $117(2)(8)$~MeV  \\
$m_s^{\overline{\mathrm{MS}}}(2\,\mathrm{GeV})$ & $103(4)(2)$~MeV  &
     $112(2)(8)$~MeV  \\
$\langle\bar\psi \psi\rangle^{\overline{\mathrm{MS}}}(2\,\mathrm{GeV})$ & 
     $-[260(6)(8) \,\mathrm{MeV}]^3$  & $-[288(5)(24) \,\mathrm{MeV}]^3$  \\
\hline
\end{tabular}
\end{center}
\caption{\label{tab-scaling}Comparison of data for the two lattices.}
\end{table}

%%%%%%%%%%%%%%%%%%%%%%%%%%%%%%%%%%%%%%%%%%%%%%%%%%%%%%%%%%%%%%%%%%%%%%%%%%%%%%

\section{Baryon spectra}

Baryon correlation functions were constructed with point sources and sinks
using standard interpolating operators.  Uncorrelated single-mass fits were
performed within fitting windows chosen on the basis of effective mass plots,
and errors were estimated by bootstrap.  For the results that follow,
baryon masses were calculated with two degenerate quarks having each of the
five lightest available masses (six on the $14^3\times 48$ lattice) and for
all available masses of the third quark.

\begin{figure}
\begin{minipage}[t]{0.48\textwidth}
\includegraphics*[width=\textwidth]{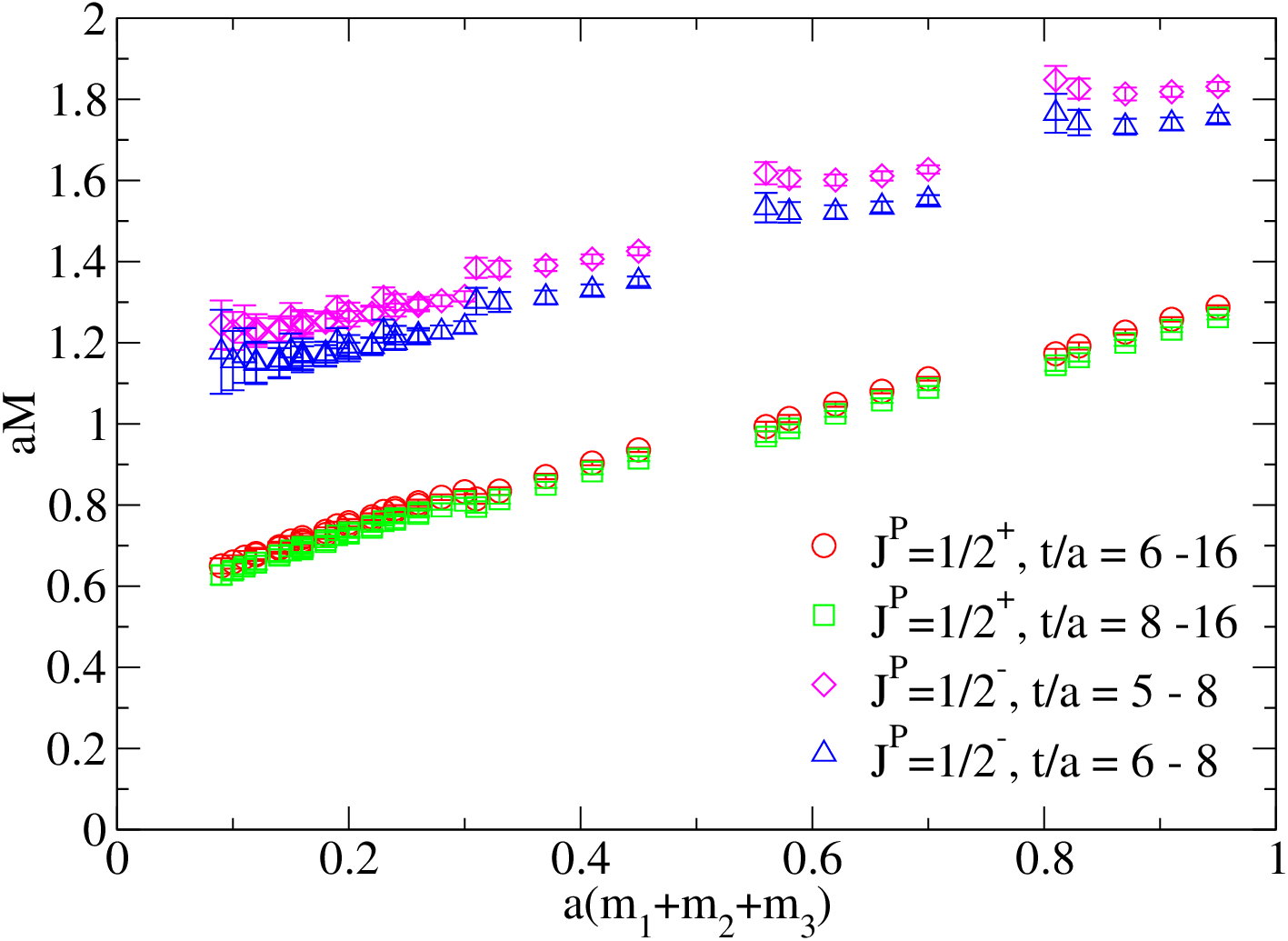}
\caption{\label{octeta}$\Lambda$-like octet masses at $\beta=6$ for two
fitting windows and quark masses $m_1=m_2$ degenerate.}
\end{minipage}
\hfill
\begin{minipage}[t]{0.48\textwidth}
\includegraphics*[width=\textwidth]{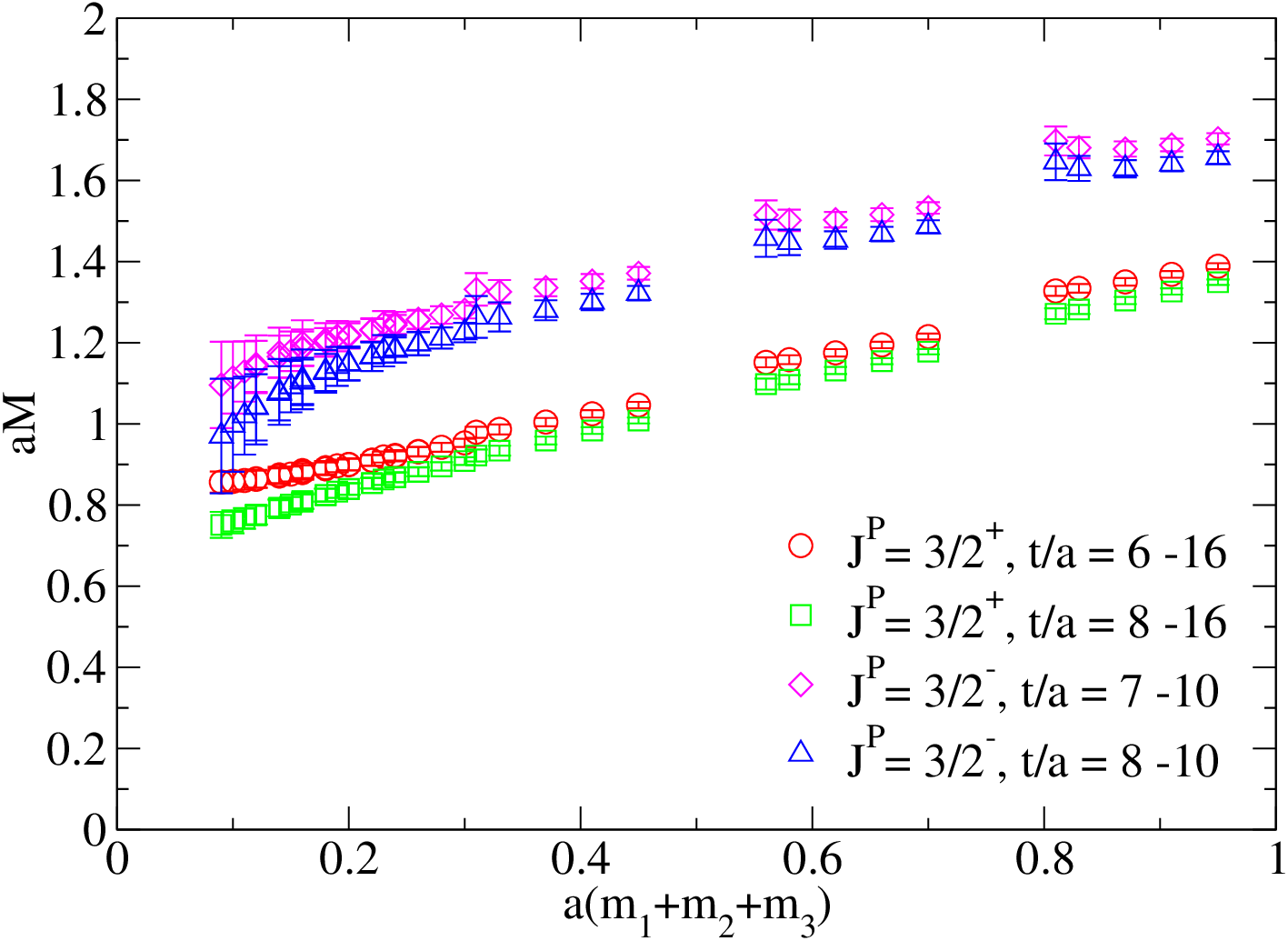}
\caption{\label{decuplet}Decuplet masses at $\beta=6$ for two fitting
windows and quark masses $m_1=m_2$ degenerate.}
\end{minipage}
\end{figure}

The positive and negative-parity octet masses are plotted in
Fig.~\ref{octeta} as a function of total quark mass.  Measurements for two
values of $t_{min}$ are shown in order to give some indication of the
dependence on fitting window.  Decuplet spectra are plotted in 
Fig.~\ref{decuplet}.

\begin{figure}
\begin{minipage}[t]{0.48\textwidth}              
\includegraphics*[width=\textwidth]{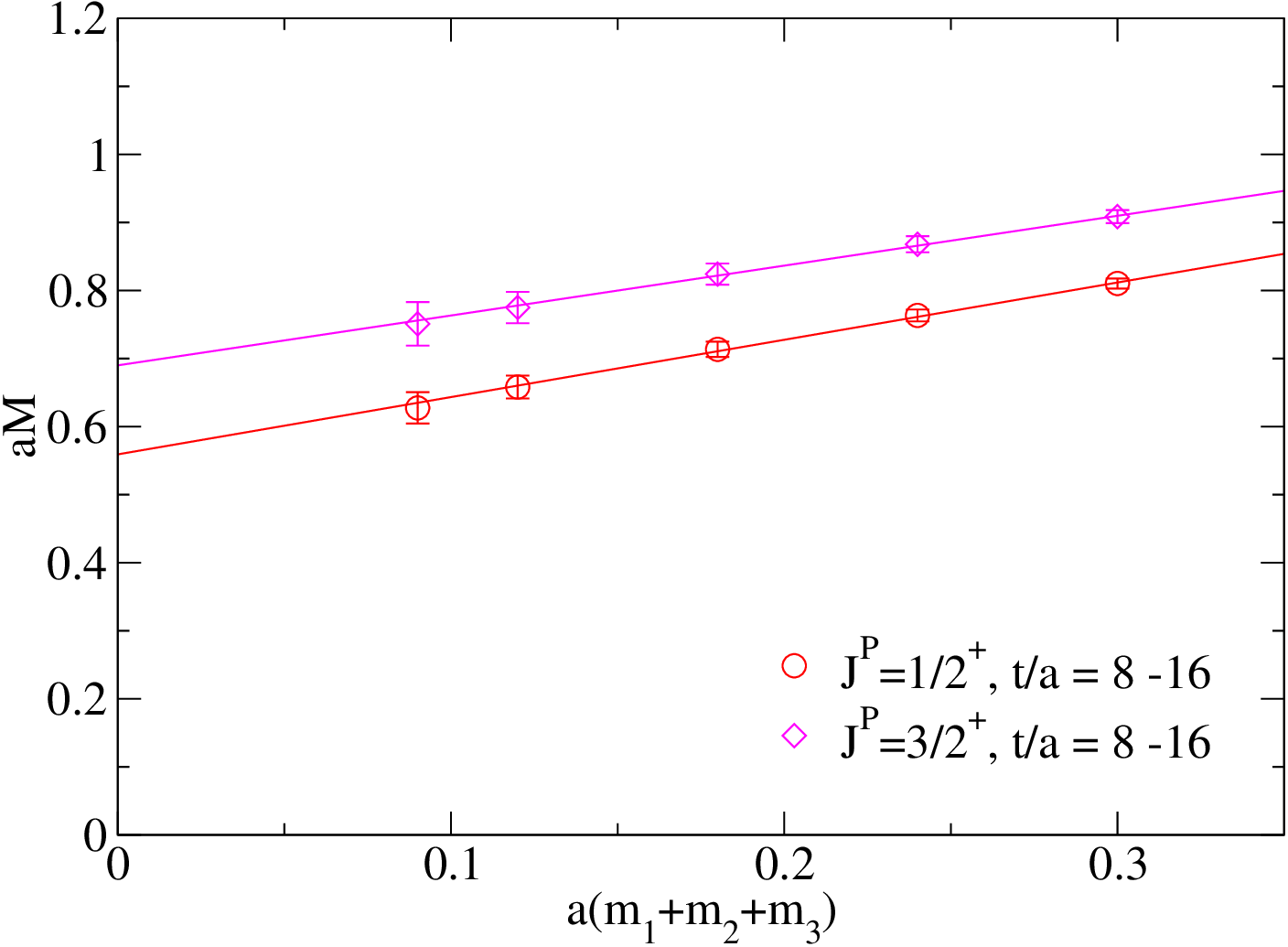} 
\caption{\label{baryon_fit}Chiral fit of baryon masses at $\beta=6$ with
three light degenerate quarks.}
\end{minipage}
\hfill
\begin{minipage}[t]{0.48\textwidth}
\includegraphics*[width=\textwidth]{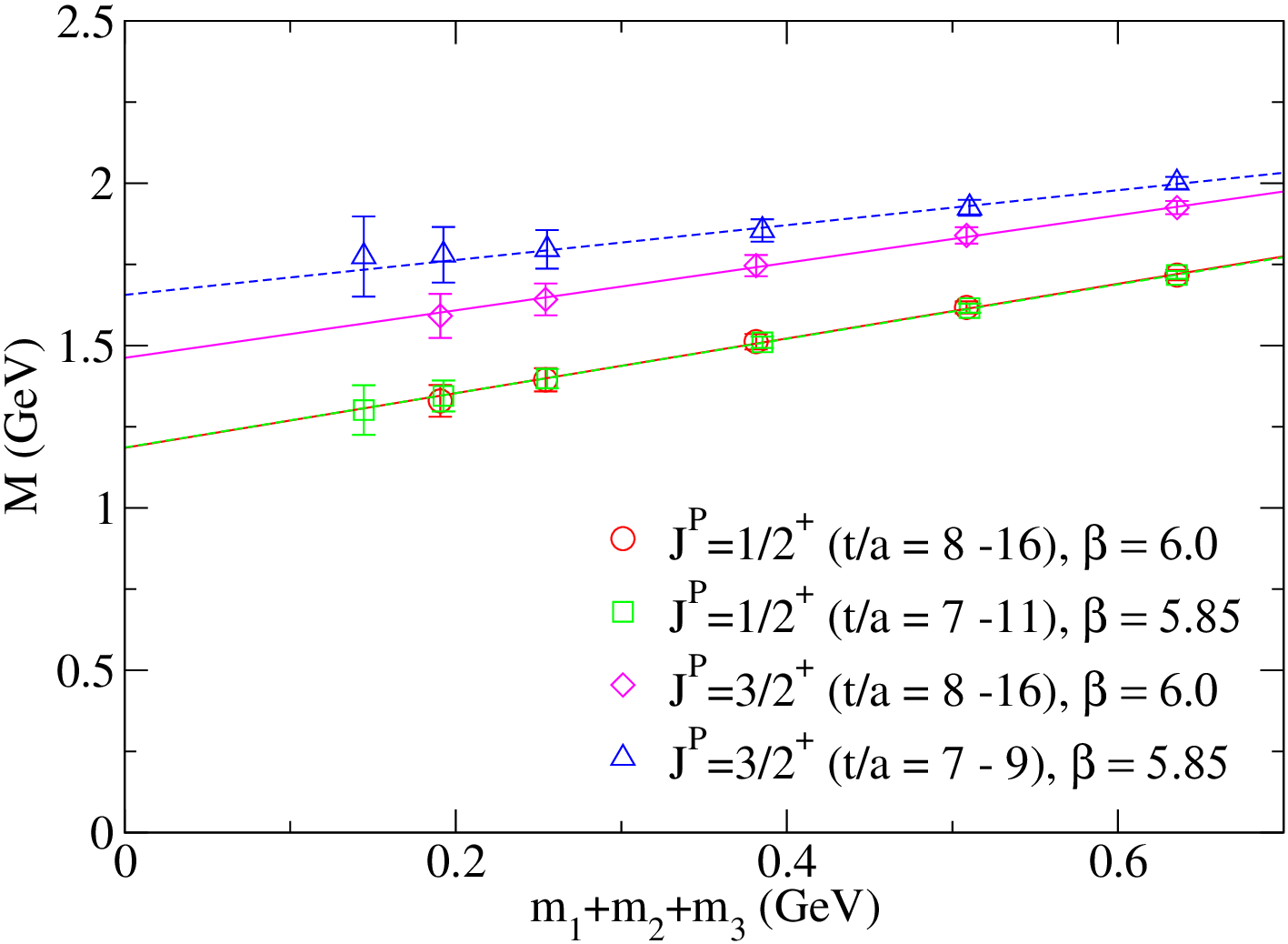} 
\caption{\label{baryon_scaling}Comparison of baryon masses with light
degenerate quarks at two values of $\beta$.}
\end{minipage}
\end{figure}

In Fig.~\ref{baryon_fit}, we plot the masses of the positive-parity states
for light degenerate quark masses $am_q=0.03,0.04,0.06,0.08,0.1$.
A linear extrapolation to the chiral limit gives $aM_8=0.559(24)$ and
$aM_{10}=0.690(32)$.  In this limit, we find $M_8/M_\rho=1.367(77)$ and
$M_{10}/M_8=1.234(78)$.

Finally, we plot in Fig.~\ref{baryon_scaling} the $J^P=\frac{1}{2}^+$ and
$J^P=\frac{3}{2}^+$ states for light degenerate quarks on both lattices where
bare masses have been rescaled by the corresponding values of $a^{-1}$ set by
the Sommer scale.  At $\beta=5.85$, we find $aM_8=0.739(28)$ and
$aM_{10}=1.032(55)$ in the chiral limit, yielding $M_8/M_\rho=1.222(61)$ and
$M_{10}/M_8=1.395(91)$.
The octet spectrum exhibits good scaling while the decuplet shows
some indication of scaling violation.  We note, however, that the decuplet
masses suffer from greater uncertainty in the choice of fitting window
(estimated to be on the same order as the statistical error),
and so the apparent lack of scaling has limited significance.

%%%%%%%%%%%%%%%%%%%%%%%%%%%%%%%%%%%%%%%%%%%%%%%%%%%%%%%%%%%%%%%%%%%%%%%%%%%%%%

\section{Diquark correlations}

The possible existence of exotic states such as the $\Theta^+$
pentaquark has given a renewed relevance to diquark models~\cite{Jaffe:2003sg}.
A study of correlations among quarks in baryons is currently in progress
which might provide additional evidence for such models.

The fact that our propagators were calculated in the Landau gauge
allows us also to measure quark-quark correlations directly and to
fit their decay in Euclidean time in terms of an effective ``diquark
mass''~\cite{Hess:1998sd}.  Of course, such a mass parameter can only be
defined in a fixed gauge and is not the mass of a physical state.
Nevertheless, it can give an indication of the relative strength of quark
bindings within diquark states. We consider correlation functions for the
diquark operators
\begin{equation}
\mathcal{O}_{c}^{s_1s_2}(x)=\epsilon_{cc_1c_2}\psi_{c_1}^{s_1}(x)\,
\psi_{c_2}^{s_2}(x)
\end{equation}
and
\begin{equation}
\mathcal{O}_{c_1c_2}^{s_1s_2}(x)=\frac{1}{\sqrt{2}}
(\psi_{c_1}^{s_1}(x)\,\psi_{c_2}^{s_2}(x)+\psi_{c_2}^{s_1}(x)\,
\psi_{c_1}^{s_2}(x)),
\end{equation}
which are a $\bf{\bar{3}}$ and $\bf{6}$ of color, respectively. Using
these operators, we form four types of diquark states: (i) color
$\bf{\bar{3}}$, spin-0, flavor $\bf{\bar{3}}$, (ii) color
$\bf{\bar{3}}$, spin-1, flavor $\bf{6}$, (iii) color $\bf{6}$, spin-0,
flavor $\bf{6}$, and (iv) color $\bf{6}$, spin-1, flavor $\bf{\bar{3}}$.

\begin{figure}
\begin{minipage}[t]{0.48\textwidth}
\includegraphics*[width=\textwidth]{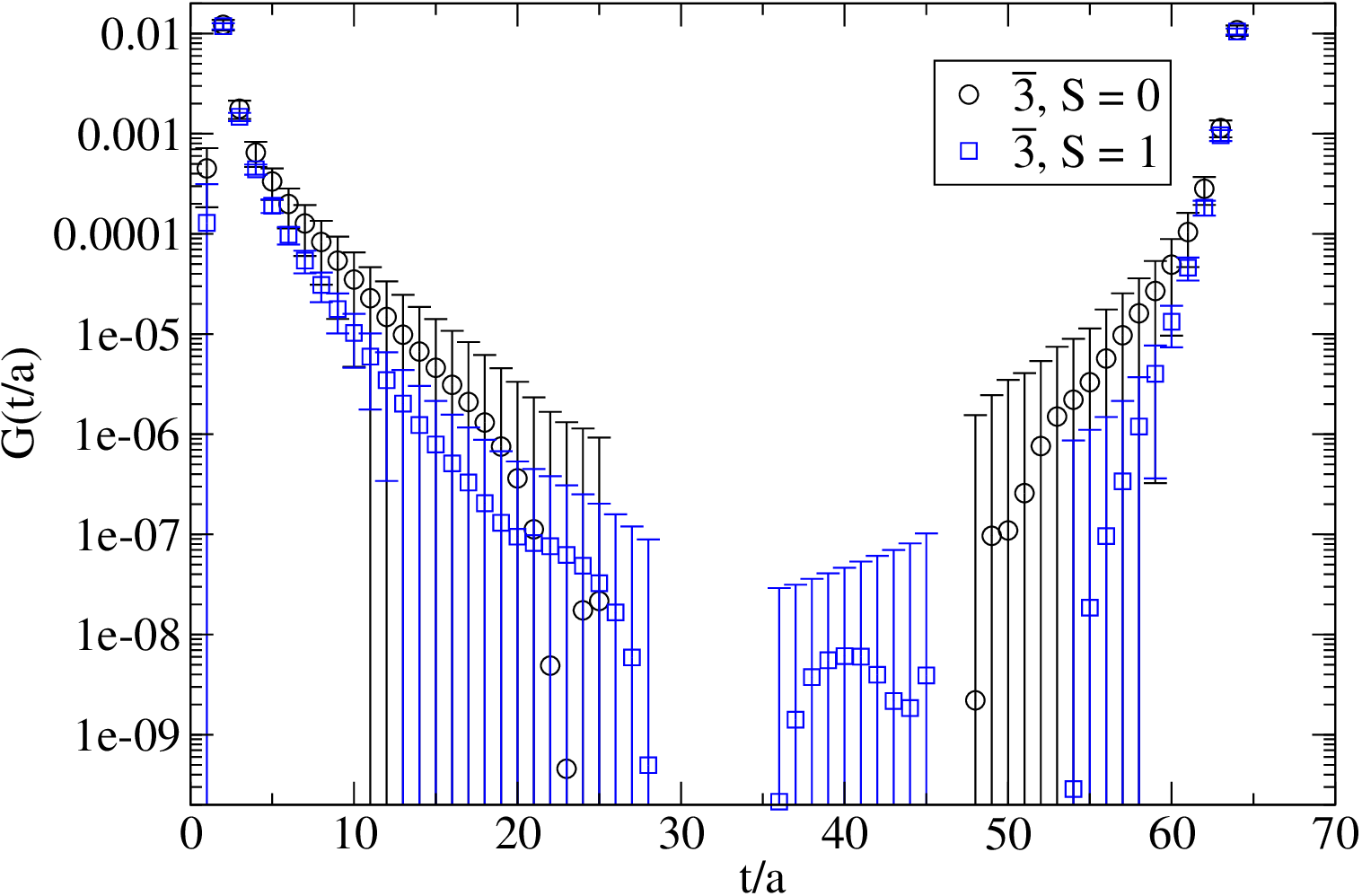} 
\caption{\label{pos-diquark}Positive-parity $\bf{\bar{3}}$ diquark correlators
for $am_1=am_2=0.03$ at $\beta=6$.}
\end{minipage}
\hfill
\begin{minipage}[t]{0.48\textwidth}
\includegraphics*[width=0.95\textwidth]{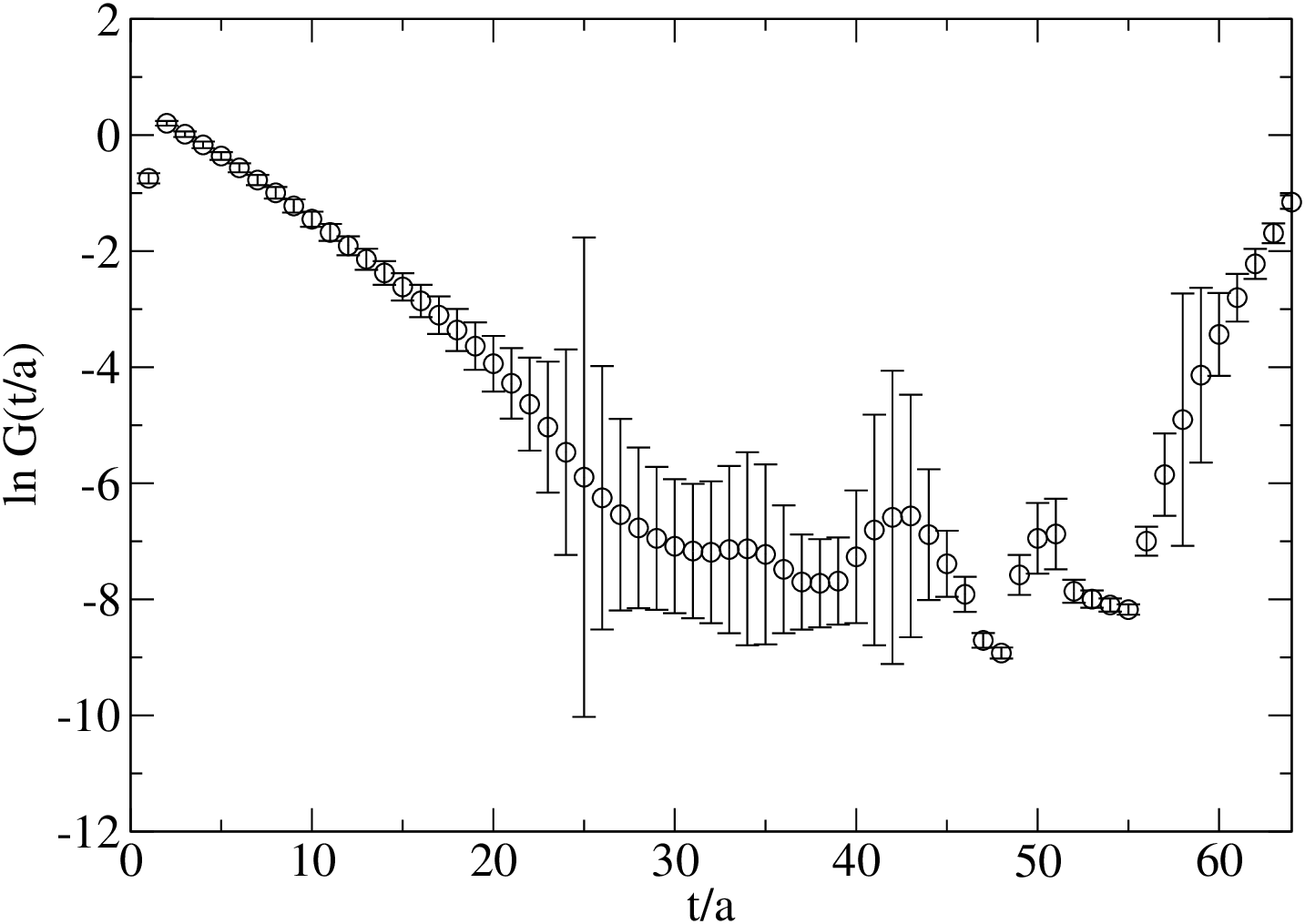} 
\caption{\label{quarkprop}Quark propagator for input quark mass $am=0.03$
at $\beta=6$.}
\end{minipage}
\end{figure}

We work in a basis where $\gamma_4$ is diagonal, and so to extract the mass
of a positive-parity state, correlators involving upper components 
(Dirac indices 1, 2) are combined with time-reversed lower-component
correlators.  Negative-parity states are constructed with one upper and one 
lower component (e.g.~1, 3).  Plotted in Fig.~\ref{pos-diquark} are
positive-parity diquark correlation functions for the $\bf{\bar{3}}$ state
with $am_1=0.03$ and $am_2=0.03$, the lightest available quark mass
combination.
For comparison we also extract constituent quark masses by performing
fits to the quark propagators.  Plotted in Fig.~\ref{quarkprop} is an
example for $am_q=0.03$, giving the constituent quark mass $aM=0.229(5)$.

\begin{figure}
\begin{minipage}[t]{0.48\textwidth}              
\includegraphics*[width=\textwidth]{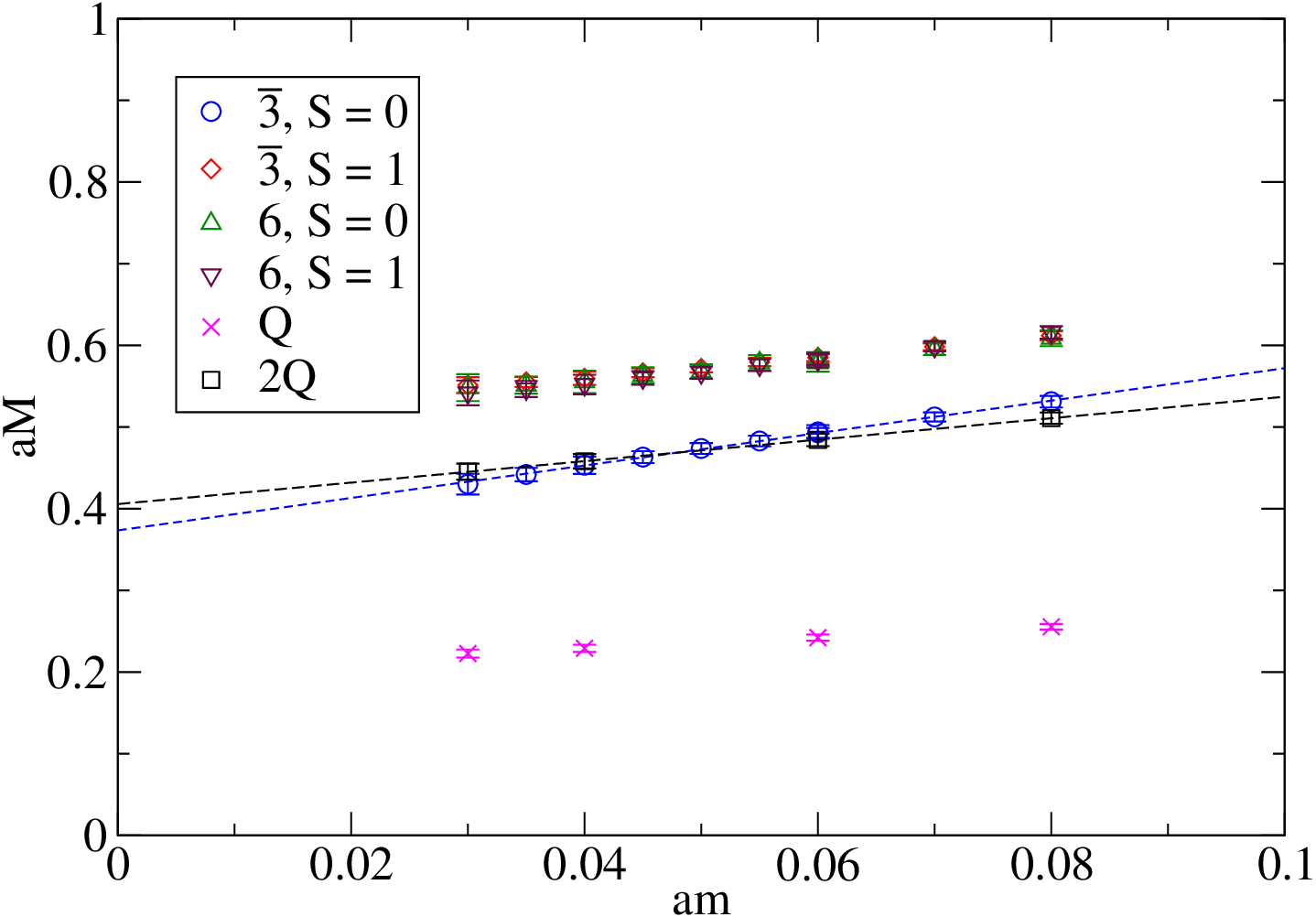} 
\caption{\label{pos-diquark-spectra}Positive-parity diquark spectra at
$\beta=6$.}
\end{minipage}
\hfill
\begin{minipage}[t]{0.48\textwidth}
\includegraphics*[width=\textwidth]{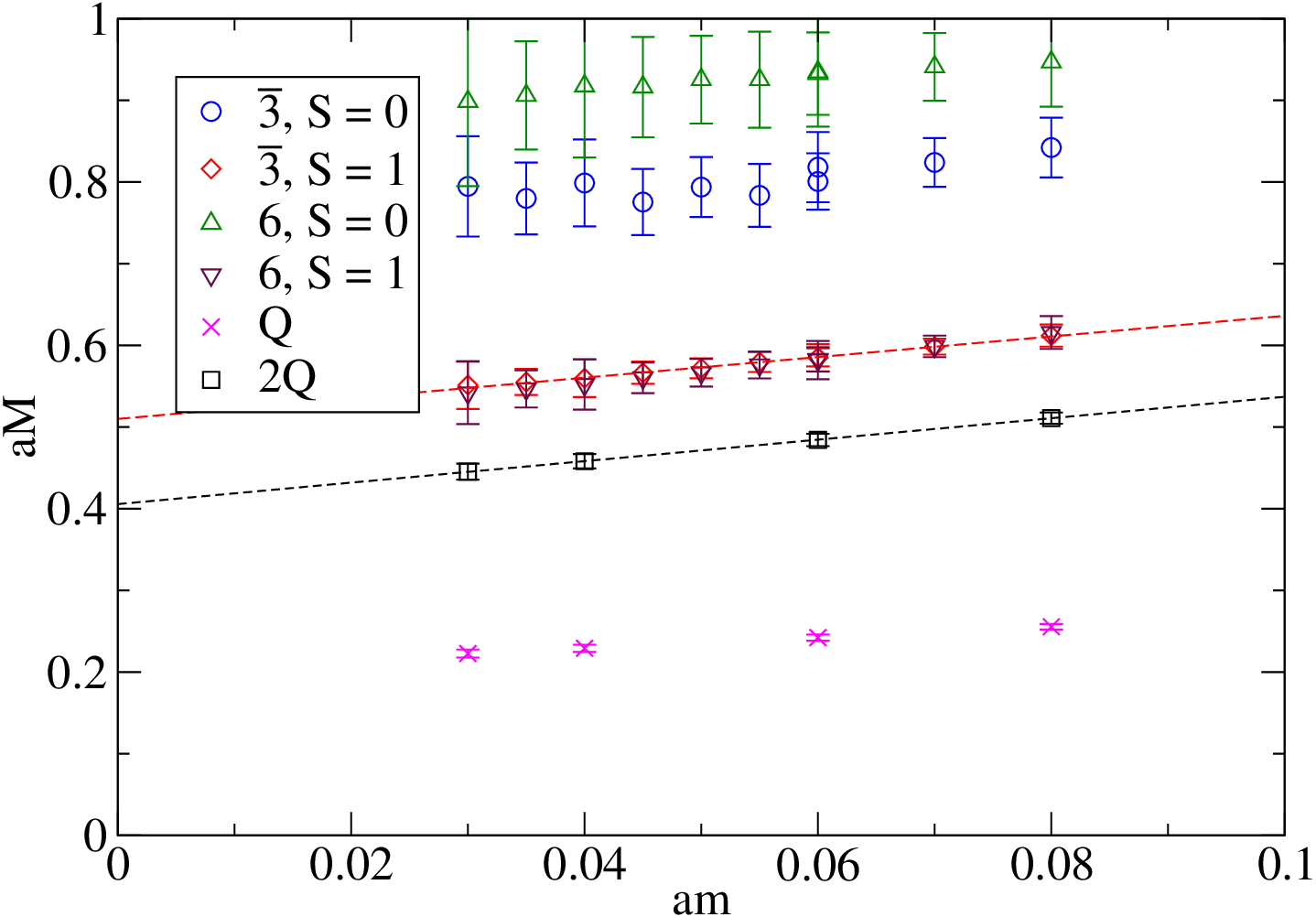} 
\caption{\label{neg-diquark-spectra}Negative-parity diquark spectra at
\mbox{$\beta=6$.}}
\end{minipage}
\end{figure}

In Figs.~\ref{pos-diquark-spectra} and~\ref{neg-diquark-spectra} we plot the
positive and negative-parity diquark spectra, as determined from fits between
$t_{min}=5a$ and $t_{max}=15a$.  Also included are plots of twice the
constituent quark mass and extrapolations to the chiral limit for these and
the lowest-lying diquark states.
It is interesting to observe that in the first plot, the
$\bf{\bar{3}}$ spin-0 diquark extrapolation is below twice the quark mass
extrapolation and that the $\bf{\bar{3}}$ spin-0 diquark state is
significantly more strongly bound than the $\bf{\bar{3}}$ spin-1 diquark.
These results are consistent with the predictions of diquark 
models~\cite{Jaffe:2003sg,Anselmino:1992vg}.
We again emphasize, however, that a much more detailed analysis must be done,
particularly of diquarks within baryons, before rigorous conclusions may be
reached.

%%%%%%%%%%%%%%%%%%%%%%%%%%%%%%%%%%%%%%%%%%%%%%%%%%%%%%%%%%%%%%%%%%%%%%%%%%%%%%

\section{Conclusions}

In this contribution, we have presented results from quenched lattice QCD
simulations using the overlap Dirac operator for meson and baryon observables,
quark masses, meson final-state wave functions, and diquark correlations.
One important result of this work is the demonstration that techniques are
mature for calculating propagators with the overlap Dirac operator on large
lattices, up to a chosen numerical precision, even with ``rough'' gauge
backgrounds.  Improved algorithms or smoothing techniques both may help to
make the calculation less computationally demanding~\cite{Durr:2005an}.

Our investigation also validates the good chiral properties of the overlap
operator and demonstrates good scaling properties between $\beta=5.85$ and
$\beta=6$, indicating that the $\beta=6$ results may already be close to the
continuum limit.  In comparisons to experiment, our results suffer from the
shortcomings of the quenched approximation.  Nevertheless, from this
investigation and others it is clear that it should be possible to use the
overlap operator in dynamical fermion simulations, at the very least with a
mixed action formulation.  Work in this direction is beginning.

\acknowledgments

This work was supported in part by US DOE grants DE-FG02-91ER40676 and
DE-AC02-98CH10866, EU HPP contract HPRN-CT-2002-00311 (EURIDICE), and grant
HPMF-CT-2001-01468.  We thank Boston University and NCSA for use of their
supercomputer facilities.

\bibliographystyle{utcaps}
\bibliography{lat05_hadron}

\end{document}